\documentclass[conference,a4paper]{APSIPA2021}
\usepackage{amsmath}
\usepackage{graphicx}
\usepackage{multirow}
\usepackage{threeparttable}
\usepackage[backend=biber,style=ieee,]{biblatex}
\usepackage{geometry}
\usepackage{fancyhdr}

\fancypagestyle{firststyle}{
  \fancyhf{}
  \fancyhead[C]{2024 Asia Pacific Signal and Information Processing Association Annual Summit and Conference (APSIPA ASC)}
}

\usepackage{amsmath,amssymb,bm,mleftright,diffcoeff}
\usepackage{siunitx}
\usepackage{graphicx}
\usepackage{booktabs}
\usepackage{makecell}
\usepackage{multirow} 
\usepackage{caption}
\usepackage{url}

\begin{document}

%\title{Evaluating L2 Utterances Intelligibility using Native Speaker Shadowing Data and Sequence to Sequence Voice Conversion}
\title{A Pilot Study of Applying Sequence-to-Sequence Voice Conversion to Evaluate the Intelligibility of L2 Speech Using a Native Speaker's Shadowings}

\author{
\authorblockA{
    Haopeng Geng, Daisuke Saito, Nobuaki Minematsu
}

\authorblockA{
Graduate School of Engineering, The University of Tokyo\\
E-mail: \{kevingenghaopeng, dsk\_saito, mine\}@gavo.t.u-tokyo.ac.jp}
}

\maketitle
\thispagestyle{firststyle}
\begin{abstract}
Utterances by L2 speakers can be unintelligible due to mispronunciation and improper prosody. In computer-aided language learning systems, textual feedback is often provided using a speech recognition engine. However, an ideal form of feedback for L2 speakers should be so fine-grained that it enables them to detect and diagnose unintelligible parts of L2 speakers’ utterances. Inspired by language teachers who correct students’ pronunciation through a voice-to-voice process, this pilot study utilizes a unique semi-parallel dataset composed of non-native speakers' (L2) reading aloud, shadowing of native speakers (L1) and their script-shadowing utterances. We explore the technical possibility of replicating the process of an L1 speaker's shadowing L2 speech using Voice Conversion techniques, to create a virtual shadower system. Experimental results demonstrate the feasibility of the VC system in simulating L1’s shadowing behavior. The output of the virtual shadower system shows a reasonable similarity to the real L1 shadowing utterances in both linguistic and acoustic aspects\footnote{Audio samples are available at: \url{https://secondtonumb.github.io/publication_demo/APSIPA_2024/index.html}}.
\end{abstract}

\section{Introduction}
Feedback from native speakers or language teachers is essential in second language acquisition (SLA), as it can improve the communicative skills of L2 speakers. With the assistance of Large Language Models (LLM) agents, it has become easier for L2 speakers to have their writing proofread with detailed suggestions. However, L2 speakers still face challenges in receiving word-by-word feedback on their speaking performance, especially when educational resources are limited.

\begin{figure}[t]
%     \vspace{-1cm}
	\centering
	\includegraphics[width=\columnwidth]{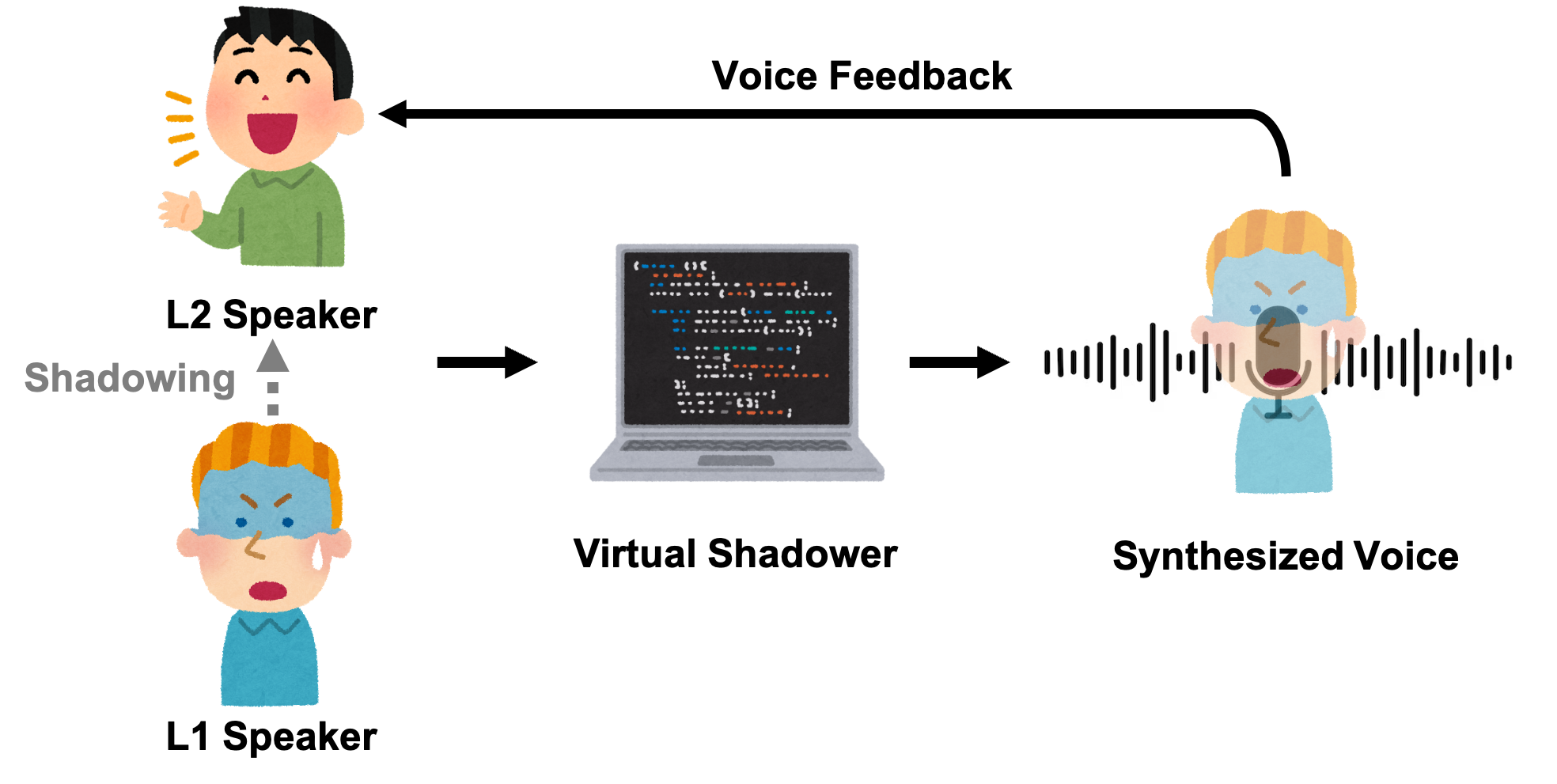}
    \vspace{-5mm}
 	\captionof{figure}{The concept of the virtual shadower, which simulates the shadowing behaviors of an L1 rater hearing a given L2 speech for the first time. Stuttering or inarticulate production of speech may occur due to listening disfluencies.}
 	\label{fig:vs_concept}
      \vspace{-5mm}
\end{figure}
\begin{figure*}[t]
	\centering
	\includegraphics[width=0.8\textwidth]{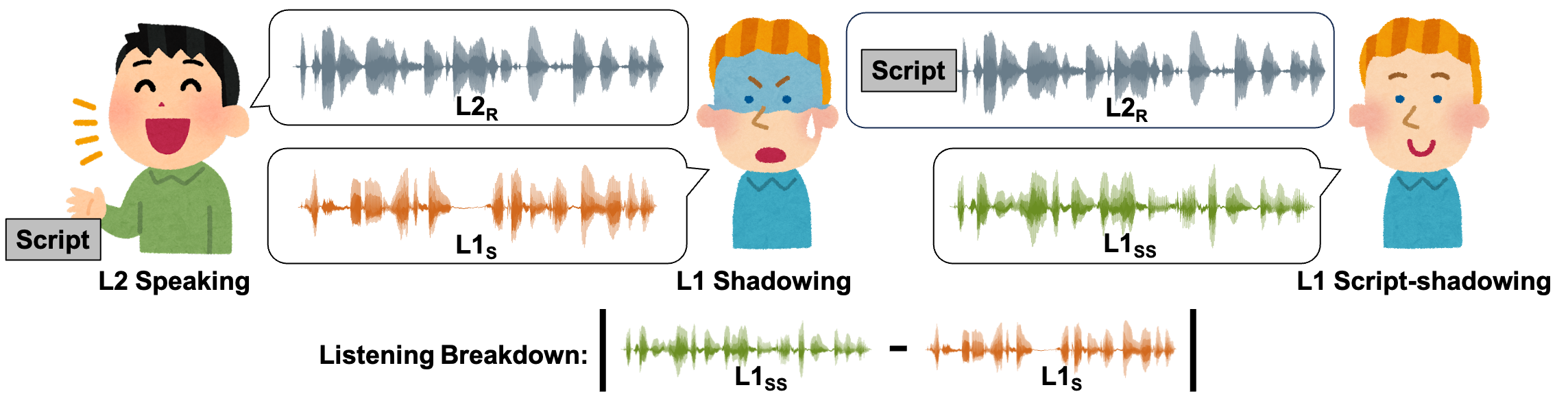}
	\captionof{figure}{The proposed shadowing technique aims to identify unintelligible parts in an L2 speaker's reading aloud utterances ($L2_{R}$). In this approach, $L1_{S1}$ represents a native speaker’s initial shadowing, while $L1_{SS}$ denotes the script-shadowing by the native speaker.
By calculating the distance between $L1_{S1}$ and $L1_{SS}$, it is possible to pinpoint the native speaker's listening breakdowns, which correspond to the unintelligible parts in the $L2_{R}$ as well.}
	\label{fig:l1shadowingl2_w_lb}
\vspace{-5mm}
\end{figure*}
In such cases, Computer-Aided Language Learning (CALL) has proven to be effective. For example, an Automatic Speech Recognition (ASR) engine can generate transcripts of an L2 speaker’s utterances, helping them understand how the system processes their speech and identify any unintelligible words. Recent advances in ASR, leveraging LLM and sophisticated feature embedding, have achieved promising performance, with Word Error Rates (WER) below 10\% for Global Englishes \cite{radford2023robust, wills_et_al:OASIcs.SLATE.2023.7, chan22b_interspeech}.

Nevertheless, previous research indicates that relying solely on ASR for speech intelligibility evaluation is not suitable for providing feedback to L2 speakers \cite{neri2001effective}. Textual feedback alone is often insufficient for L2 speakers to identify unintelligible segments, especially in spontaneous situations, as the system has no access to the ground truth reference which is considered to exist only in the speaker's mind. Moreover, recent ASR systems are specifically designed to predict, or even over-speculate, what speakers actually said, which is contrary to the purpose of CALL systems. \cite{weng2020joint,ma2023can}. Besides, research indicates that L2 speakers often overestimate their speech intelligibility \cite{Derwing_Munro_1997, munro1995foreign} and replaying their utterances is generally less effective than anticipated in helping them recognize unintelligible parts of their speech. In the worst case, replaying L2 speakers' speech may fossilize their pronunciation.

% To our knowledge, no ASR systems are designed to mis-recognize L2 speech to simulate real human listeners. 

How can we expose those unintelligible parts in L2 speech? Native speaker shadowing, wherein an L1 speaker repeats an L2 utterance with as short a delay as possible while listening, is proposed as an advanced method for evaluating L2 speech intelligibility \cite{inoue18_interspeech, lin2020shadowability, zhu2021multi}. In this approach, the L1 speaker immediately repeats what they hear in the given L2 speech using their own accent. Any stuttering or delivery of unrelated words, known as shadowing disfluencies, indicates where listening breakdowns occur. Consequently, shadowing by the L1 speaker highlights the problematic parts of the L2 speech, enabling L2 speakers to know where their speech may be unclear or difficult to understand.
However, it is impractical to provide a real shadower for every L2 speaker. In this study, as shown in Figure~\ref{fig:vs_concept}, we aim to develop a virtual shadower system that simulates the process of an L1 speaker shadowing L2 speech (L1-shadowing-L2), thereby offering more comprehensible and constructive feedback for L2 speakers.

% Since the shadower can generate the corresponding context almost simultaneously, speakers are not restricted to pre-determined material, allowing for the assessment of spontaneous utterances. 

%This approach involves delivering the results through human-uttered speech, where mispronounced segments are emphasized by the L1 speaker.

Our research contributions are as follows:
\begin{itemize}
\item We examine the similarity between an L1 shadower's behavior and Voice Conversion (VC) alignment, utilizing this similarity to construct a virtual L1-shadowing-L2 system, which simulates an L1 speaker's shadowing L2 speech. To the best of our knowledge, this is the first work to apply VC to the L2 intelligibility task.
    
\item We examine the feasibility of using L1 shadowing speech data, a form of semi-parallel data, to develop a VC model. Experimental evaluations on both linguistic and acoustic aspects demonstrate reasonable feasibility in constructing an L1-shadowing-L2 system with Seq2Seq VC.

\end{itemize}

\section{Research Background}
\subsection{Native speaker shadowing}

Fine-grained annotation of the intelligibility of L2 speech is such a challenging task that it requires specialist knowledge in phonetics for phoneme-level labeling. To address this problem, \cite{lin2020shadowability} proposed a two-stage reverse form \footnote{In the field of SLA, shadowing typically involves learners repeating L1 speech to enhance their listening skills. However, in our study, learners are shadowed by L1 raters. Thus, we refer to our method as reverse shadowing.}  of shadowing to identify unintelligible parts in L2 utterances.
As shown in Figure \ref{fig:l1shadowingl2_w_lb}, an L1 speaker first shadows a given L2 utterance alone ($L1_{S1}$). During this process, unintelligible parts result in stuttering or inarticulate production.
Following $L1_{S1}$, the L1 speaker performs script-shadowing ($L1_{SS}$), where the L2 script is presented visually along with the L2 utterance. By comparing $L1_{S1}$ and $L1_{SS}$ using Dynamic Time Warping (DTW), we obtain sequential data on broken articulation—shadowing disfluencies that indicate listening disfluencies—based on the distance between the two.

Previous research has shown that alignment based on phonetic posteriorgrams (PPG) performs better than acoustic features such as MFCC or Mel-Cepstrum in assessing pronunciation quality \cite{yue17_interspeech}. Additionally, PPG-based DTW between $L1_{S1}$ and $L1_{SS}$ can derive word, syllable, and phoneme-level annotations of intelligibility,  as also shown in \cite{zhu2021multi}.

%when given an L2 utterance,
% where an L1 speaker’s initial shadowing (S1) and script-shadowing (SS) utterances are compared.

\subsection{Seq2Seq voice conversion}
Conventional VC aims to change non-/para-linguistic features while preserving the linguistic content of input speech. However, with the advent of end-to-end architectures and self-supervised speech representations, recent studies on Seq2Seq VC enable a more robust mapping of sophisticated latent features between source and target voices. This includes tasks such as Speaking Style Conversion \cite{maimon-adi-2023-speaking}, Voice Emotion Conversion \cite{zhou_emotionVC}, and Foreign Accent Conversion \cite{zhao_fac_journal, zhao_ppg}.

In \cite{huang20i_interspeech}, the authors proposed the Voice Transformer Network (VTN) that utilizes a Transformer-based Seq2Seq model and employs mel-spectrograms as input features \cite{attention}. In \cite{fac-evaluate}, the authors modified the input acoustic feature to linguistic representation, resulting in a significant reduction of accentedness. To address the issues of inaccurate duration prediction and repetitive artifacts caused by the auto-regressive nature of transformer models, a non-autoregressive model was proposed using a conformer structure to mitigate these problems \cite{hayashi2021non,ren2020fastspeech}. Recent works have further improved upon this by using a Monotonic Alignment Search (MAS) and joint vocoder training, outperforming non-autoregressive models in duration and prosody \cite{okamoto2023e2e}. Furthermore, a more robust Automatic Alignment Search (AAS) method, derived from a Text-to-Speech (TTS) system, has been developed to address challenges in low-resource settings \cite{huang2023aas, shih2021rad}.

In the following sections, these recent advancements in Seq2Seq VC are effectively adapted to implement our purposed L1-shadowing-L2 system.

\section{Experimental settings}
\subsection{Representing listening breakdowns with attention alignment failure}
\label{ses: dtwtoatt}
\begin{figure}[t]
	\centering
	\includegraphics[width=\columnwidth]{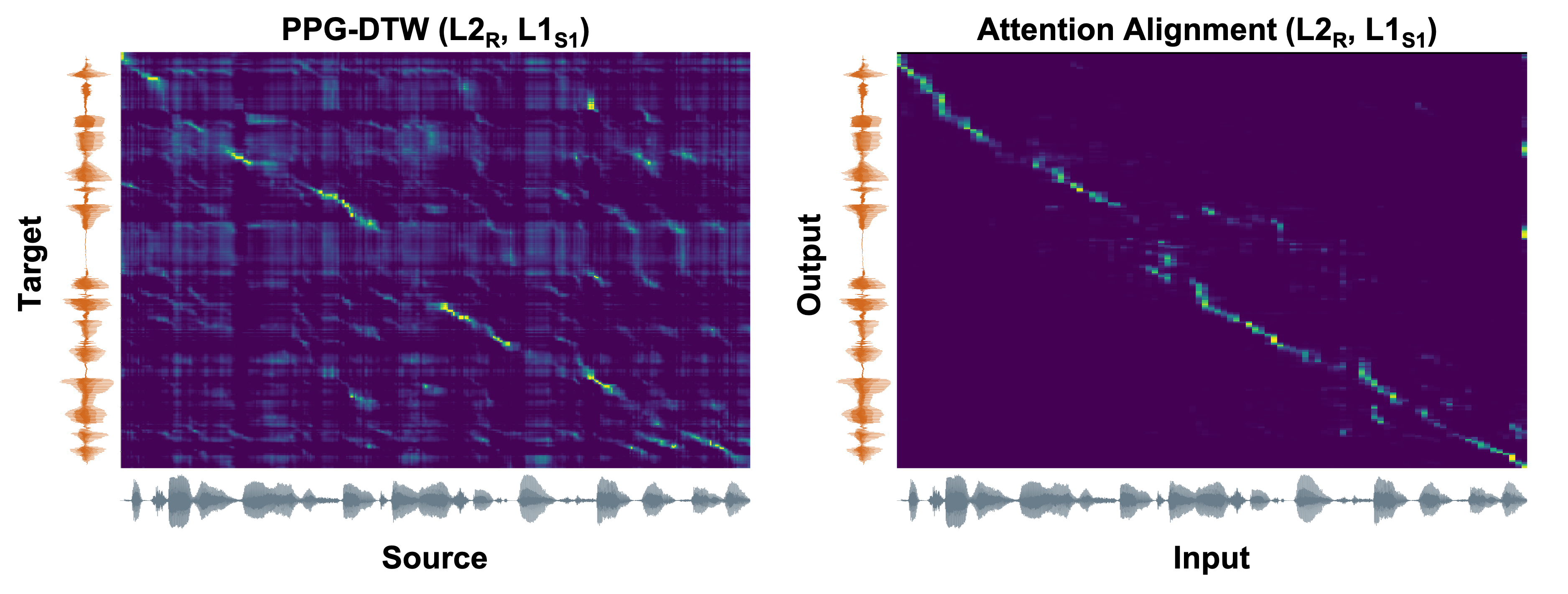}
	\captionof{figure}{The similarity of attention alignment to PPG-DTW is illustrated. 
The left figure shows the attention alignment observed in the inference phase of converting $L2_{R}$ to $L1_{S1}$, while the right figure shows the PPG-DTW path between $L2_{R}$ and $L1_{S1}$. Both figures exhibit a prominent diagonal contour.}
	\label{fig:att_dtw_simi}
     \vspace{-5mm}
\end{figure}

Shadowing is a voice-to-voice process in which the listener’s instantaneous identification of words is crucial. However, smooth shadowing does not always occur, as listening breakdowns can happen when an L1 listener encounters unintelligible segments while shadowing L2 speech \cite{hamada2014effectiveness}. To explore the potential of VC to build virtual shadowers, Figure~\ref{fig:att_dtw_simi} shows our pilot experiment on Seq2Seq VC using semi-parallel data, $L2_{R}$ and $L1_{S1}$.  We observe that the decoding path in the $L2_{R}$-$L1_{S1}$ VC system closely resembles the PPG-DTW editing path. This inspires us to explore whether a VC system can be extended to explicitly represent L2 unintelligibility. If alignment failure (weak or faded-out attention) indicates listening breakdowns, we suggest that such a virtual shadower can be developed to resemble a real human L1 shadower.

\subsection{Experimental setups}

\subsubsection{VC models}
To ensure that the VC system’s alignment failure is attributed to data mismatch rather than the model’s limited alignment capabilities (as discussed in Section~\ref{ses: dtwtoatt}), we evaluated two VC models in this study. Alongside the baseline Transformer-based model, VTN \cite{vtn_journal}, we tested a robust alignment method, AAS, proposed in AAS-VC \cite{huang2023aas}.

\subsubsection{Data preparation}

In this study, we utilized a reverse form of  shadowing dataset mentioned in \cite{yue17_interspeech, zhu2021multi}.  With different degrees of accent, 225 Japanese English speakers' reading-aloud utterances $L2_{R}$ were collected. We then recruited one male native English speaker who shadowed and script-shadowed all these recordings as $L1_{S1}$ and $L1_{SS}$ respectively. 
Although each recording session lasted for more than 30 seconds, to avoid potential alignment failure due to lengthy input, we trimmed the original recordings sentence by sentence using forced alignment. 
This process resulted in 2,695 triplets of \{$L2_{R}$, $L1_{S1}$, $L1_{SS}$\}. Each dataset comprised an average of 3.9 hours of valid phonation, with the average duration of each sentence being 5.0 seconds. 300 utterances were selected as test sets.

\subsubsection{Source-target selection}
To determine the optimal setting for an L1 virtual shadower simulator, we designed three different source-target pairings:
\begin{itemize}
    \item $L2_R$-$L1_{S1}$: This is the most straightforward approach for building a virtual shadower, as the training target is $L1_{S1}$ itself. However, it is also the most technically challenging setting, as both the linguistic context and speaker identity differ.
    \item $L2_R$-$L1_{SS}$: This is the typical setting for parallel-data VC. Since the source and target are intended to speak the same content, we expect this setting to reveal the pronunciation distance between L2 and L1.
    \item $L1_{SS}$-$L1_{S1}$: This unique training setting of the reverse shadowing dataset focuses on the listening breakdowns of the L1 shadower. We expect that this setting will reveal the disfluencies exhibited during the shadowing behavior of L1.
\end{itemize}

\subsubsection{Feature embedding and pretrained models for decoding}
\label{sec:pretrain_models}

Speaker-independent features enable stable convergence in our many-to-one VC setting. For feature embedding, we used the original PPG-like bottleneck feature extractor designed by \cite{9428161}. The framewise dimension of latent feature is 144, and the features were normalized to zero mean and unit variance for embedding.

As for the PPG-to-Spec decoding, we trained a single-speaker PPG-to-Spec decoder following the implementation described in \cite{s3prl-vc-journal}, where utterances were collected from the given L1 speaker. 

For the vocoder, we employed Parallel-WaveGAN \cite{yamamoto2020parallel}, which was trained on joint $L1_{S1}$ and $L1_{SS}$. The input Mel spectrogram has 80 frequency bins with a hop size of 256, and the target sample rate is 16 kHz.

Note that for VTN, both the PPG-to-Spec decoder and vocoder were utilized in the decoding phase. For AAS-VC, we followed the original implementation to map the source PPG to the target Mel-spectrogram, which  generates the target’s prosodic features better as previous study discussed \cite{huang2023aas}.

\section{Experimental Evaluation}

\begin{table*}
  \centering
\caption{CER/WER and S1-CER/WER results evaluated on original recordings, PPG-VC, VTN and AAS-VC outputs.}
  \begin{tabular}{@{}lllllllll@{}}
  \toprule
 & \multirow{2}{*}{Model}& \multicolumn{2}{c}{Training}& \multirow{2}{*}{Testset}     & \multirow{2}{*}{CER$\downarrow$}& \multirow{2}{*}{WER$\downarrow$}& \multirow{2}{*}{S1-CER $\downarrow$}&\multirow{2}{*}{S1-WER$\downarrow$}\\
  &   & Source   & Target    && & & & \\
  \midrule
$\dagger$&   -   & -  & -  &$L1_{S1}$   & 6.83  & 13.71 & - & - \\
$\dagger$&     & -  & -  &$L1_{SS}$   & 0.85  & 3.60  & - & - \\
$\dagger$&      & -  & -  &$L2_{R}$        & 10.04 & 20.91 & - & - \\
  \midrule    
$\ast$&   PPG-VC \cite{liu2021any}  & $L1_{spk}$      & $L1_{spk}$     &$L1_{S1}$   & 8.00  & 15.90 & - & - \\
$\ast$&    &      $L1_{spk}$      &   $L1_{spk}$     &$L1_{SS}$   & 1.90  & 5.70  & - & - \\
$\ast$$\ast$&    & $L2_{spk}$      & $L2_{spk}$     &$L2_{R}$        & 21.67 & 38.83 & - & - \\
$\ast$$\ast$$\ast$&    & $L2_{spk}$      & $L1_{spk}$     &$L2_{R}$        & 21.81 & 39.23 & - & - \\
  \midrule
$\ddagger$  & VTN \cite{huang20i_interspeech}     & $L2_{R}$      & $L1_{S1}$    &$L2_{R}$   & 22.64 & 37.29  & 22.41 & 36.15\\
$\ddagger$  &   & $L2_{R}$      & $L1_{SS}$    &$L2_{R}$   & {\bf 19.47} & {\bf 34.53} & 19.28 & 33.53 \\
$\ddagger$  &    & $L1_{SS}$ & $L1_{S1}$    &$L2_{R}$   & 32.21 & 51.24 & 32.08 & 50.56\\
\midrule
$\ddagger$  & AAS-VC \cite{huang2023aas}    &  $L2_{R}$ & $L1_{S1}$  &$L2_{R}$   & 21.60 & 38.68 & 21.53 & 37.98\\
$\ddagger$  &    &  $L2_{R}$ & $L1_{SS}$  &$L2_{R}$   & {\bf 20.44} & {\bf 36.74} & 19.98 & 35.97\\
$\ddagger$  &     &  $L1_{SS}$ & $L1_{S1}$  &$L2_{R}$   & 26.49 & 46.76 & 26.49 & 46.50\\
\bottomrule
\end{tabular}

\label{tab:vs_cer_wer}
\end{table*}

\begin{table*}[h!]
\centering
\caption{Mel Cepstral Distortion (MCD), F0 Root Mean Square Error (F0RMSE), F0 Correlation (F0CORR), and Absolute Duration Difference (DURR) on PPG-VC, VTN and AAS-VC. PPG-VC converts multiple L2 speaker identities into the target L1 speaker to minimize the influence of speaker identity on acoustic features.}
\begin{tabular}{@{}lllllllll@{}}
\toprule
 \multirow{2}{*}{Model}  & \multicolumn{2}{c}{Training}& \multirow{2}{*}{Testset}& \multirow{2}{*}{Reference}& \multirow{2}{*}{MCD$\downarrow$} &\multirow{2}{*}{F0RMSE$\downarrow$} & \multirow{2}{*}{F0CORR$\uparrow$ }&\multirow{2}{*}{DURR$\downarrow$}\\

 & Source   & Target    & && & & & \\
\midrule
     -       &     -     & -   & $L2_{R}$     &$L1_{S1}$& 12.84 & 89.65 & 0.084 & 0.337 \\
     &     -     & -   & $ L1_{SS}$  &$L1_{S1}$& 6.62 & 35.24  & 0.385 & 0.350 \\
\midrule
PPG-VC    & $L2_{R}$& $L1_{spk}$& $L2_{R}$&$L1_{S1}$& 8.94 & 47.27 & 0.127 & 0.339 \\

          &                   $L2_{R}$& $L1_{spk}$& $L2_{R}$&$L1_{SS}$& 8.97 & 47.03  & 0.117  & 0.361 \\
\midrule
VTN       &                   $L2_{R}$& $L1_{S1}$   & $L2_{R}$&$L1_{S1}$& 7.53 & 40.35  & 0.239  & 0.552 \\
          &                   $L2_{R}$& $L1_{SS}$   & $L2_{R}$&$L1_{S1}$& 7.47 & 39.76  & 0.253  & 0.575 \\
\midrule
AAS-VC    &                   $L2_{R}$& $L1_{S1}$   & $L2_{R}$&$L1_{S1}$& 7.39 & 38.76  & 0.240  & 0.515 \\
          &                   $L2_{R}$& $L1_{SS}$   & $L2_{R}$&$L1_{S1}$& 7.34 & 37.78  & 0.259  & 0.401 \\

\bottomrule
\end{tabular}
\label{tab:mcd_f0rmse}
\end{table*}

% \midrule
%          & $L2$              & $L1_{S1}$   & 12.84 & 89.65 & 0.084 & 0.337 \\
%          & $L2$              & $L1_{SS}$   & 12.85 & 89.17 & 0.073 & 0.361 \\
%          & $L1_{SS}$         & $L1_{S1}$   & 6.62 & 35.24  & 0.385 & 0.350 \\
\subsection{Metrics for evaluation}

To objectively evaluate our virtual shadower, we focused on two key aspects. First, the linguistic content of the virtual shadower should closely match that of the L1 human shadower. Second, the acoustic similarity of the virtual shadower should resemble that of the L1 shadower. In this pilot study, we assessed the linguistic output of our virtual shadower only holistically. For the acoustic aspect, we calculated the segmental and prosodic similarities between the virtual shadower and the L1 shadower. Before presenting the evaluation results for these two aspects, the following section will show the ASR results for $L1_{S1}$, $L1_{SS}$, and $L2_R$, which are necessary for subsequent discussions.

\subsection{Ablating effects of vocoding module}
We first applied ASR to all testing data using a reputable and advanced ASR system\footnote{\url{https://huggingface.co/facebook/wav2vec2-large-960h-lv60-self}} \cite{wav2vec2}. In this study,  the ASR results reflect the expressions of the L1 shadower during the shadowing sessions (S1 and SS) and the L2 learner during the recording session (R). Inevitably, the ASR result may include errors, which are quantified using the CER/WER by comparing the results with the reference text used during the recording of $L2_R$. As shown in $\dagger$ of Table \ref{tab:vs_cer_wer},  the differences in CER/WER values between $L1_{S1}$ and $L1_{SS}$ reveal the linguistic variations introduced by shadowing. The $L2_{R}$ utterances, characterized by a strong accent, consequently display even higher CER/WER compared to the L1 speech. 

To eliminate propagation errors potentially caused by VC decoders, we performed a framewise PPG-based VC that only converts the speaker identity based on the implementation in \cite{liu2021any} for analysis synthesis (ANA-SYN). The PPG decoder and vocoder described in Section~\ref{sec:pretrain_models} were utilized. The CER/WER of the ANA-SYN speech is presented in $\ast$, $\ast$$\ast$ and $\ast$$\ast$$\ast$ of Table~\ref{tab:vs_cer_wer}.  $\ast$ and $\ast$$\ast$ show the VC results in which the identity of the input speaker was maintained, while $\ast$$\ast$$\ast$ shows the conversion of $L2_{spk}$ to $L1_{spk}$. The minimal variations in CER/WER for $L1_{S1}$ and $L1_{SS}$ suggest that the embedding features do not distort linguistic information. However, a larger degradation in CER/WER for ANA-SYN $L2_{R}$ is noted in $\ast\ast$, which is considered to be due to the inevitable noises present in $L2_{R}$. Note that $L2_{R}$ recordings were collected separately in various environments using personal devices, making it difficult to ensure standardized acoustic quality. Nevertheless, the very small difference in CER/WER after speaker normalization to $L1_{spk}$, shown in $\ast$$\ast$$\ast$, indicates that these recording disturbances do not significantly affect the decoder's performance.

\subsection{Linguistic similarity of virtual shadower}
\label{sec:ling}
Since the proposed virtual shadower aims to construct potentially corrupted speech generated by the L1 shadower, the output is different from what the L2 learner actually intends, but expects to be more similar to what the shadower said. While the conventional WER is calculated as follows:

\begin{equation}
\text{WER}= \frac{S + I + D}{N_{R}},
\end{equation}
we introduce S1-WER to evaluate the linguistic similarity between the conversion results  and $L1_{S1}$ based on their ASR results.

\begin{equation}
\text{S1-WER}= \frac{S + I + D}{N_{S1}},
\end{equation}
where $S$, $I$, and $D$ are substitution, insertion, and deletion errors of the converted results, respectively. $N_{R}$ represents the word count of L2 learners' handcrafted scripts, while $N_{S1}$ is the word count of the recognition hypothesis of $L1_{S1}$ conducted by the mentioned ASR system. Lower S1-WER will represent higher linguistic similarity between the converted speech and the shadowing speech. S1-CER is the character-based error rate, calculated in a similar way to S1-WER.

As shown in $\ddagger$ of Table~\ref{tab:vs_cer_wer}, compared to $\ast$$\ast$$\ast$, the reduction of WER on $L2_{R}$-$L1_{S1}$ and $L2_{R}$-$L1_{SS}$ demonstrates the accent reduction capability of Seq2Seq VC.
The best CER/WER are achieved using the $L2_{R}$-$L1_{SS}$ training pair, as the L1 shadower fluently repeated the content intended by the L2 learner by viewing the learner's script, which is a beneficial condition for VC training. For the $L2_{R}$-$L1_{S1}$  which is the most straightforward approach to achieve virtual shadowing by the L1 speaker, the results are reasonable, as the semi-parallel setting brings high barriers for feature mapping from a mispronounced segment to the correct segment. In the $L1_{SS}$-$L1_{S1}$ setting, the relatively high CER/WER indicates that the model struggles to accurately capture the unintelligibility of the L2 speakers. This is likely because $L2_{R}$’s features are considered out-of-domain.

Regarding S1-CER/WER, although the differences are not significant, all proposed models have lower or equal S1-CER/WER values compared to CER/WER. This indicates that Seq2Seq VC systems have the potential to effectively map semi-parallel linguistic features, such as $L2_{R}$-$L1_{S1}$ in this study. Particularly, $L1_{SS}$ targeting models outperform all the other training settings, indicating the possibility of building a virtual shadowing system utilizing parallel data only. 

\subsection{Acoustic similarity of virtual shadower}

Table~\ref{tab:mcd_f0rmse} illustrates the acoustic similarity between $L1_{S1}$ and the converted speech. Compared to the $L1_{spk}$-normalized $L2_{R}$ conducted by PPG-VC, the better metrics in MCD and F0 demonstrate the prosody reconstruction ability of the proposed virtual shadower. In particular, AAS-VC, benefiting from its superior alignment ability, shows better prosody similarity, especially in F0 and duration.

Interestingly, although the model was not trained with $L1_{S1}$, the results showed that the output of $L1_{SS}$-targeting model closely resembles the acoustic similarity of the actual $L1_{S1}$.  This indicates the potential of using parallel data directly to establish a virtual shadowing system, which supports the conclusion mentioned in Section~\ref{sec:ling}.

\section{Conclusions and future works}
\subsection{Conclusions}
In this study, we first proposed the concept of a virtual shadower, which simulates the behavior of an L1 listener who instantly repeats what an L2 speaker utters while listening to it. By reflecting the L1 listener’s immediate understanding, the virtual shadower aims to highlight linguistic ambiguities in L2 speech that the speaker may not notice. Experiments conducted on two Seq2Seq VC systems with various training data settings from the {$L2_R$, $L1_{S1}$ and $L1_{SS}$} triplet demonstrated the potential of parallel and semi-parallel training from $L2_R$ to $L1_{S1}$/$L1_{SS}$. 

\subsection{Future works}
\subsubsection{Generalization ability}
Shadowing behavior varies among listeners with different language backgrounds. This study involved only Japanese learners of English and a single rater. Future experiments will involve participants who speak different languages and multiple raters to explore the capability and adaptability of Seq2Seq VC as more flexible and general virtual shadowers.

% \subsubsection{Quantified correlation analysis between DTW and attention based alignment}
% While we observe similarities between the DTW editing path and the attention decoding path on the paired $L2_{R}$ and $L1_{S1}$, further quantifiable evaluation is necessary to provide stronger evidence. For instance, using a loss function derived from PPG-DTW ($L1_{S1}, L1_{SS}$) is considered a more promising and interpretable approach.

\subsubsection{Integrating $L1_{SS}$-$L1_{S1}$ information for virtual shadowing}
In this study, we acknowledge and examine the challenge of directly mapping $L2_{R}$-$L1_{S1}$ as they are both linguistically and acoustically different.
However, the promising result with $L2_{R}$-$L1_{SS}$  encourage us to combine information derived from $L1_{SS}$-$L1_{S1}$. We believe the distance between $L1_{SS}$-$L1_{S1}$ has not been fully utilized yet. For instance, using a loss function derived from PPG-DTW ($L1_{S1}, L1_{SS}$) is considered a more promising and interpretable approach. Additionally, exploring the cascading of $L2_{R}$-$L1_{SS}$ with $L1_{SS}$-$L1_{S1}$ could yield valuable insights to convert $L2_R$ to $L1_{S1}$.

\subsubsection{Self-supervised speech representations with prosodic feature considered}
Although linguistically-rich embedding like PPG perform well in foreign accent conversion tasks, PPG lacks prosodic features such as intonation and rhythm, which often affect listeners' comprehension. Self-supervised speech representations have shown great potential in reconstructing unintelligible speech, such as electrolaryngeal or dysarthric speech \cite{lester_EL}. We will explore the feasibility of using latent features from models like HuBERT \cite{hsu2021hubert}  and WavLM \cite{chen2022wavlm}, which are known to include both linguistic and  prosodic features.

\subsubsection{Pedagogical Assessment of virtual shadower}
Our virtual shadower offers feedback by pinpointing segments of learners’ speech that are unintelligible to listeners. Unlike conventional feedback, which is typically provided as scores or written comments, our feedback is delivered as immediate vocal response. This approach offers more realistic feedback, which we expect will enhance learners’ motivation more effectively than traditional methods.

\printbibliography

\end{document}